\documentclass[aps,prd,10pt,preprint,superscriptaddress,onecolumn,eqsecnum,nofootinbib]{revtex4-1}

\usepackage[utf8]{inputenc}
\usepackage[T1]{fontenc}

\usepackage{graphicx}
\usepackage{epsfig}

\usepackage{amsmath,amssymb,amsfonts}
\usepackage{dsfont}
\usepackage{mathrsfs}
\usepackage{amsthm}
\usepackage{mathtools}
\usepackage{bm}

\usepackage{xcolor}
\usepackage[normalem]{ulem}

\usepackage{siunitx}

\usepackage{booktabs}
\usepackage{hhline}

\usepackage{blindtext}

\usepackage{orcidlink}

\usepackage{hyperref}

\hypersetup{
  colorlinks=true,
  urlcolor=blue,
  linkcolor=magenta,
  citecolor=blue,
  filecolor=blue,
  pdftitle={Black holes and wormholes in Einstein--Cartan--Maxwell gravity with a conformal scalar field},
  pdfauthor={Luis Avil\'es, Omar Valdivia, Rodolfo V\'eliz, Carlos Vera},
  pdfsubject={Einstein--Cartan gravity, torsion, black holes, wormholes},
  pdfkeywords={torsion, Einstein--Cartan gravity, conformal scalar field, black holes, wormholes}
}

\newcommand{\diff}[1]{\text{d}#1}
\newcommand{\Diff}[1]{\text{D}#1}

\begin{document}
	
\title{Torsional black holes and wormholes in Einstein–Cartan–Maxwell gravity with a conformal scalar field}

	\author{Luis Avil\'es}
	\email{luaviles@unap.cl}
	\affiliation{Instituto de Ciencias Exactas y Naturales, Universidad Arturo Prat, Playa Brava 3256, 1111346, Iquique, Chile}
	\affiliation{Facultad de Ciencias, Universidad Arturo Prat, Avenida Arturo Prat Chac\'on 2120, 1110939, Iquique, Chile}
	\author{Omar Valdivia}
	\email{ovaldivi@unap.cl}
	\affiliation{Instituto de Ciencias Exactas y Naturales, Universidad Arturo Prat, Playa Brava 3256, 1111346, Iquique, Chile}
	\affiliation{Facultad de Ciencias, Universidad Arturo Prat, Avenida Arturo Prat Chac\'on 2120, 1110939, Iquique, Chile}
	
	\author{Rodolfo V\'eliz}
	\email{rvelizm@estudiantesunap.cl}
	\affiliation{Instituto de Ciencias Exactas y Naturales, Universidad Arturo Prat, Playa Brava 3256, 1111346, Iquique, Chile}
	\affiliation{Facultad de Ciencias, Universidad Arturo Prat, Avenida Arturo Prat Chac\'on 2120, 1110939, Iquique, Chile}

\author{Carlos Vera}
	\email{carlosvera02@unap.cl}
	\affiliation{Instituto de Ciencias Exactas y Naturales, Universidad Arturo Prat, Playa Brava 3256, 1111346, Iquique, Chile}
	\affiliation{Facultad de Ciencias, Universidad Arturo Prat, Avenida Arturo Prat Chac\'on 2120, 1110939, Iquique, Chile}

\begin{abstract}

%We investigate a gravitational theory formulated in the first-order formalism, where the metric and connection are treated as independent fields. Within this framework, we introduce a one-parameter extension of conformal transformations that allows for the conformal coupling of a scalar field while naturally generating nontrivial torsion. By coupling the model to Einstein–Cartan gravity and to an electromagnetic field, we construct exact static solutions in asymptotically flat and asymptotically locally AdS regimes. Depending on the values of the integration constants and the deformation parameter, the theory admits black hole and wormhole configurations supported by a scalar field that remains regular throughout the spacetime. In particular, we obtain a regular black hole solution in the absence of electric charge, as well as topological black holes in the asymptotically AdS case whose very existence is tied to the presence of electric charge. In the vanishing-torsion limit, the solutions reduce to previously known configurations in purely metric theories. These results provide new scenarios for the study of regular spacetimes and gravitational systems with torsion. 

We formulate a one-parameter extension of Weyl transformations in first-order gravity and show that it defines a conformally coupled scalar sector with dynamical torsion. The construction reduces to the standard torsionless conformal coupling in the limit $\lambda\to 1$. In the corresponding Einstein--Cartan--Maxwell theory, we derive exact static solutions in asymptotically flat and asymptotically locally AdS spacetimes. These solutions describe scalar-dressed black holes, regular black holes, and traversable wormholes, depending on the values of the integration constants and of the deformation parameter. We show that torsion can regularize the scalar field and, for suitable branches, also improve the geometric singularity structure. In the AdS sector, the existence of topological black holes requires a nonvanishing electric charge. These results provide new exact examples of regular and torsional configurations in four-dimensional gravity.

\end{abstract}

\maketitle

\section{Introduction}

Black holes remain among the most compelling objects in theoretical physics, serving as natural laboratories to test the foundations of gravitation. Within Einstein's General Relativity, stationary black hole solutions coupled to minimally interacting matter are subject to strong uniqueness results: in the Einstein–Maxwell theory, they are fully characterized by their mass, charge, and angular momentum, in accordance with the celebrated no-hair theorems~\cite{Israel:1967wq,Israel:1967za,Carter:1971zc,Ruffini:1971bza,Bekenstein:1995un,Chrusciel:2012jk}. This rigidity, however, relies on the assumption of minimal coupling and on the purely Riemannian nature of the underlying geometry. Once these assumptions are relaxed, novel families of black hole solutions with nontrivial ``hair'' can emerge, thereby broadening the scope of gravitational phenomena.

A paradigmatic example arises in the context of conformally coupled scalar fields. Unlike minimally coupled scalars—which in asymptotically flat spacetimes typically lead to naked singularities—conformal couplings admit exact solutions in which the scalar field contributes nontrivially to the geometry. The pioneering works of Bekenstein~\cite{Bekenstein:1974sf} and, independently, Bocharova, Bronnikov, and Melnikov~\cite{Bocharova:1970skc} established the existence of such configurations, although the scalar field diverges at the event horizon. Initially regarded as pathological, subsequent analyses showed that this divergence does not necessarily invalidate the black hole character of the solution~\cite{Bekenstein:1975ts}. These developments stimulated further research, including black hole solutions in the presence of a negative cosmological constant, where the scalar singularity can be hidden behind the horizon, rendering the field effectively regular for external observers~\cite{Martinez:2002ru,Martinez:2005di}. More generally, extensions have been explored in asymptotically AdS spacetimes ~\cite{Caldarelli:2013gqa,Anabalon:2012tu,Ayon-Beato:2015ada}, including configurations coupled to different matter sectors \cite{Anabalon:2009qt,Aviles:2026fsf}, higher dimensions \cite{Hassaine:2007py,Hendi:2012as,Charmousis:2012dw,Giribet:2014fla,BravoGaete:2013acu}, and within modified gravity frameworks such as Horndeski theories and conformal higher-curvature models \cite{Anabalon:2013oea,Cisterna:2014nua,Saridakis:2016mjd,Chernicoff:2016jsu,Ayon-Beato:2024vph,Corral:2025npd}. A common feature of these scenarios is that nonminimal scalar couplings provide natural mechanisms to circumvent the limitations imposed by classical no-hair results.

A complementary and conceptually rich direction arises within the first-order formulation of gravity, in which the metric and connection are treated as independent dynamical variables, thereby allowing for nonvanishing torsion. In this framework, nonminimally coupled scalar fields can act as natural sources of torsion~\cite{Barrientos:2017utp}, an effect that appears in low-energy effective string models, in cosmological scenarios involving accelerated expansion~\cite{Izaurieta:2020xpk}, and in axion-like couplings associated with the Nieh–Yan density (see, for instance,~\cite{Zwiebach:1985uq,Toloza:2013wi,Castillo-Felisola:2015ema}). From the perspective of black hole physics, it has recently been shown in~\cite{Aviles:2024muk} that spacetime torsion can significantly modify the structure of the solutions, as it can render nonminimally coupled scalar fields regular everywhere, an outcome that is generally not possible in purely Riemannian geometries.

On the other hand, several studies have explored the role of torsion in the construction of wormhole solutions within the Einstein–Cartan framework and related metric-affine theories \cite{Anchordoqui:1996ah,GarciadeAndrade:2000pb,Mehdizadeh:2018smu,Krechet:2018kmg,Ziaie:2019dmq,Soni:2023tso}. In particular, torsion can contribute effectively to the gravitational dynamics in a manner that relaxes the need for exotic matter sources or modifies the effective stress–energy tensor supporting the wormhole geometry \cite{Bronnikov:2015pha,Jawad:2016zwj,DiGrezia:2017daq,Sarkar:2024efj}. Wormhole configurations supported by scalar fields, spin fluids, or generalized matter couplings have been reported in the literature~\cite{Farfan:2011ic,Mehdizadeh:2017dhb,Mehdizadeh:2017tcf,Errehymy:2023rsm,Yousaf:2025vcg,Sarkar:2025iiz}, suggesting that torsion can play a central role in ensuring the regularity of the geometry and in facilitating the formation of a wormhole throat. These findings indicate that spacetimes endowed with nontrivial torsion provide a natural framework for investigating traversable wormholes and their physical properties.

Motivated by these developments, in this work we further investigate this possibility by considering a gravitational theory formulated in the first-order formalism, in which torsion arises dynamically through a conformally coupled scalar field. Within this framework, we construct exact static solutions describing wormhole geometries supported by a regular scalar field, thereby extending previous results in Einstein–Cartan gravity.

More specifically, we consider four-dimensional Einstein–Cartan gravity coupled simultaneously to a Maxwell field and to a conformally invariant scalar field. We show that this theory admits charged black hole geometries in which the scalar field remains regular throughout the spacetime. Furthermore, we find that the existence of asymptotically AdS black hole solutions is intrinsically linked to the presence of electric charge, whereas in the asymptotically flat regime and in the absence of charge the theory admits regular black hole and wormhole configurations. These solutions provide new counterexamples to the traditional no-hair paradigm, thereby extending the class of scalar-dressed black holes in four dimensions and underscoring the role of torsion as a fundamental geometrical mechanism responsible for preserving scalar field regularity.
Altogether, our results show that a one-parameter generalization of conformal symmetry in Riemann--Cartan geometry provides a unified mechanism for generating exact four-dimensional solutions with nontrivial torsion, including regular black holes and traversable wormholes, while continuously recovering the corresponding purely metric configurations in the limit $\lambda\to 1$.
We emphasize that the regularity of the scalar field does not automatically imply the regularity of the spacetime geometry. In what follows, we distinguish between branches with a regular scalar field and those that also possess finite curvature and torsional invariants, reserving the term regular black hole for the latter case.

The manuscript is organized as follows. In Sect.~II, we review Einstein–Cartan gravity and introduce the conformally coupled scalar field within the first-order formalism, together with the inclusion of the Maxwell field in the theory. In Sect.~III, we investigate exact static solutions describing black hole and wormhole configurations with nontrivial torsion, considering different asymptotic regimes of the spacetime. Finally, in Sect.~IV, we conclude with a discussion of our main results and outline possible directions for future research.

\section{Theoretical Framework}
\subsection{Einstein–Cartan gravity} 
The Einstein-Cartan framework extends Riemannian geometry by treating the metric and the affine connection as independent variables, thereby allowing for a nonvanishing spacetime torsion tensor. 

Let $M_4$ be a four-dimensional manifold. The orthonormal coframe (vielbein) one-form $e^{a}=e^{a}{}_{\mu}\mathrm{d}x^{\mu}$ encode the metric through
\begin{equation}
g_{\mu\nu}=\eta_{ab}e^{a}{}_{\mu}e^{b}{}_{\nu},
\end{equation}
where $\eta_{ab}=\mathrm{diag}(-1,1,1,1)$ is the Minkowski metric. The affine structure is described by the spin connection one-form $\omega^{ab}=\omega^{ab}{}_{\mu}\mathrm{d}x^{\mu}$, which acts as the gauge field of local Lorentz transformations. With these variables, the torsion and curvature two-forms are defined as
\begin{align}
R^{ab}&=\mathrm{d}\omega^{ab}+\omega^{a}{}_{c}\wedge \omega^{cb}\,, \label{riem}\\
T^{a}&=\mathrm{d}e^{a}+\omega^{a}{}_{b}\wedge e^b\,.
\end{align}
These satisfy the Bianchi identities
$\Diff{R^{ab}=0}$ and $\Diff{T^a}=R^{a}{}_b\wedge e^b$.
A key observation is that the spin connection can be decomposed into a torsion-free and torsion-full part, namely the Levi-Civita connection $\mathring{\omega}^{ab}$ and the contortion one-form $\kappa^{ab}$ such that
\begin{equation}
R^{ab}=\mathring{R}^{ab}+\mathring{\mathrm{D}}\kappa^{ab}+\kappa^{a}{}_{c}\wedge\kappa^{cb}\,,
\end{equation}
where $\mathring{R}^{ab}$ and $\mathring{\mathrm{D}}$ are the Riemann curvature and the covariant derivative built from the Levi-Civita connection, respectively\footnote{We use circles on top of quantities which are torsion-free}.

The dynamical content of Einstein-Cartan gravity is obtained through the functional variation of the four-dimensional action
\begin{equation}\label{ecar}
\mathrm{I}[e,\omega,\Psi]=\frac{1}{4\kappa}\int_{M_4}\epsilon_{abcd}\left(R^{ab}-\frac{\Lambda}{6}e^{a}\wedge e^{b}\right)\wedge e^{c}\wedge e^{d}+\mathrm{I_{m}}[e,\omega,\Psi]\,,
\end{equation}
where $\kappa=8\pi G$, $\Lambda$ is the cosmological constant, and $\Psi$ denotes generic matter fields. The equations of motion are obtained by independent variations of the vielbein and the spin connection
\begin{align}
\frac{1}{2}\epsilon_{abcd}R^{ab}\wedge e^{c}-\frac{\Lambda}{6}\epsilon_{abcd}e^{a}\wedge e^{b}\wedge e^{c}=\kappa \, *\tau_{d}\,,\\
\epsilon_{abcd}T^{c}\wedge e^{d}=\kappa \,  *\sigma_{ab}\,,
\end{align}
where $*$ stands for the Hodge dual operator which maps $p$ into $(d-p)-$forms. The functional variation of $\mathrm{I_m}[e,\omega,\Psi]$ defines the stress-energy one-form $\tau^{d}$ and the spin-energy density one-form $\sigma^{ab}$ as
\begin{align}
\delta_{e}\mathcal{L}_{\mathrm{m}}&=- *\tau_{d}\wedge \delta e^{d} \,,\\
\delta_{\omega}\mathcal{L}_{\mathrm{m}}&=-\frac{1}{2}\delta\omega^{ab}\wedge *\,\sigma_{ab} .
\end{align}
    \subsection{Conformal Riemann-Cartan structure}
Conformal symmetry plays a central role in the coupling of scalar fields to gravity. In the context of Riemann–Cartan geometry, where the vielbein and spin connection are independent variables, the notion of a conformal transformation must be generalized beyond the metric sector.

Let us first recall that a  standard Weyl rescaling for the metric is defined by
\begin{equation}
g_{\mu\nu}\mapsto \exp(2\alpha)g_{\mu\nu}
\end{equation}
for a local scalar parameter $\alpha=\alpha(x)$. In terms of the vielbein one-forms, this transformation is realized as
\begin{equation}
e^{a}\mapsto \exp(\alpha)e^{a}
\end{equation}
In a torsion-free Riemannian setting, the spin connection transforms accordingly so as to preserve metric compatibility. However, once torsion is introduced, the vielbein and contorsion are independent degrees of freedom, and there is no unique prescription for how to perform Weyl rescaling in the affine sector of geometry. To address this, one may parameterize a family of conformal transformations as 
\begin{subequations}\label{CT}
    \begin{eqnarray}
	e^a \mapsto& \bar{e}^{a}  & = \exp(\alpha)
    e^{a}\label{cf1}\,,\\
	\omega^{ab} \mapsto& \bar{\omega}^{ab}  &= \omega^{ab} + \lambda \,\theta^{ab}\label{cf2}\,, \\
    \kappa^{ab} \mapsto& \bar{\kappa}^{ab} &= \kappa^{ab}+(\lambda-1)\theta^{ab}\,,
\end{eqnarray}
\end{subequations}
where $\lambda$ is a dimensionless deformation parameter interpolating between nonequivalent conformal prescriptions, and $\theta^{ab}$ is a one-form constructed from the rescaling function $\alpha(x)$
\begin{equation}
\theta^{ab}=e^{a}\xi^{b}-\xi^{a}e^{b},
\end{equation}
with
\begin{equation}
\xi^{a}=\mathrm{I}^{a}\mathrm{d}\alpha\,.
\end{equation}
 Here, $\mathrm{I}^{a}$ denotes an interior product defined with respect to the background vielbein which is constructed in terms of the Hodge dual operator.
 Two particular cases are noteworthy. For $\lambda=1$, the transformation reduces to the canonical Weyl scaling: the full spin connection transforms as in the torsionless case, while the contorsion remains invariant. For $\lambda=0$, the spin connection remains fixed, and the entire effect of the rescaling is absorbed by the contorsion \cite{Nieh:1981xk}. Intermediate values of $\lambda$ interpolate between these two definitions \cite{Chakrabarty:2018ybk,Izaurieta:2020kuy}.
 This construction ensures that the torsionless condition is preserved only in the canonical case $\lambda=1$. More generally, under these generalized conformal transformations, the torsion and curvature two-forms transform as
 \begin{align}
  T^{a}\mapsto  \exp({\alpha}) T^{a}+(\lambda-1){e}^{a}\wedge \mathrm{d}(\exp(\alpha))\,,\\
 R^{ab}\mapsto R^{ab}+\lambda \mathrm{D}\theta^{ab}+\lambda^{2}\theta^{a}{}_{c}\wedge\theta^{cb}\,.
 \end{align}
 This generalized conformal structure provides the geometric foundation for coupling scalar fields in a Weyl-invariant manner to Einstein–Cartan gravity. Crucially, it illustrates that in a Riemann–Cartan geometry the choice of transformation law for the contorsion is not unique, and that the resulting dynamics may retain conformal invariance while modifying the role of torsion in the coupled system.
    \subsection{Maxwell sector in Riemann–Cartan geometry}
An interesting observation is that electromagnetic fields couple to the metric structure of spacetime but do not interact directly with torsion. In the language of differential forms, the Maxwell field is described by a one-form potential $A=A_{\mu}\mathrm{d}x^{\mu}$, with the associated field strength two-form $F=\mathrm{d}A$. Since $F$ is defined as the exterior derivative of the gauge potential, it is manifestly invariant under the abelian gauge transformation $A\mapsto A+\mathrm{d}\chi$ with $\chi(x)$ an arbitrary function of spacetime coordinates. This definition does not involve the spin connection and consequently, the presence of torsion does not alter the basic structure of Maxwell's theory.

The corresponding action in a four-dimensional Riemann–Cartan spacetime takes the standard form
\begin{equation}\label{MaxAct}
\mathrm{I_A}[A,e]=-\frac{1}{8\pi}\int_{M_4}F\wedge*F\,,
\end{equation}
where again $*$ is the Hodge dual constructed in terms of the metric encoded in the vielbein. Variation of the action with respect to $A$ yields the field equations $\mathrm{d}*F=0$ while the Bianchi identity $\mathrm{d}F=0$ is automatically satisfied. These are precisely the Maxwell equations in curved spacetime, with the distinction that the metric structure arises from the independent vielbein in the Riemann–Cartan setting.

Thus, the Maxwell sector is insensitive to the torsional degrees of freedom of the connection, and its dynamics remain formally identical to the Riemannian case. Nevertheless, torsion may influence the electromagnetic field indirectly through its effect on the geometry and the definition of conserved charges in spacetimes with nontrivial asymptotics.

    \subsection{Total action}
The full dynamical system we consider consists of three sectors: the Einstein–Cartan gravitational action, the conformally invariant scalar field, and the Maxwell field. Each is formulated in the first-order formalism, where the vielbein $e^{a}$ and spin connection $\omega^{ab}$ are treated as independent variables.

The gravitational sector is governed by the Einstein-Cartan action $\mathrm{I_{EC}}[e,\omega]$, Eq.\eqref{ecar}. The matter sector $\mathrm{I_{m}[e,\omega,\phi]}$ is composed of  a conformally coupled field $\phi$ to the geometry. Moreover, the electromagnetic sector is described by a gauge invariant Maxwell action $\mathrm{I_{A}}[A,e]$, Eq.\eqref{MaxAct}. Putting these contributions together, the total action we consider $\mathrm{I}[e,\omega,\phi,A]=\mathrm{I_{EC}}[e,\omega]+\mathrm{I_{A}}[A,e]+\mathrm{I_{m}[e,\omega,\phi]}$ which is given by

\begin{align}
\mathrm{I}[e,\omega,\phi,A]&=\int_{\mathcal{M}_4} \Bigg[\frac{1}{4\kappa}\epsilon_{abcd}\left(R^{ab}-\frac{\Lambda}{6} e^{a}\wedge e^{b}\right)\wedge e^{c}\wedge
	e^{d}-\frac{1}{8\pi}F \wedge *  F \notag \\
 \label{exact1}
	&-\frac{1}{24}\epsilon_{abcd}\left( \phi^2R^{ab} + 
	\bigg\{ \lambda\left[ 1- \frac{\lambda}{2} \right]  Z^{2}+ V(\phi) \bigg\} e^{a}\wedge e^{b} +4\lambda \phi Z^{a} T^{b}\right)\wedge e^{c}\wedge e^{d}\Bigg]\,.%  \nonumber\\
%	&-\mathrm{d}\left(\frac{\lambda}{12}\,\phi\,\epsilon_{abcd}\,Z^a\,e^b\wedge e^{c}\wedge e^{d}\right) ,   
\end{align}
where $Z^{a}=\mathrm{I}^{a}\mathrm{d}\phi=e^{a}{}_{\mu}\nabla^{\mu}\phi$. The matter sector is invariant under the conformal transformations~\eqref{CT}, provided that the scalar field transforms as $\bar{\phi}=\exp(-\alpha)\phi$ and that the scalar potential either vanishes or takes the conformal form $V(\phi)=\phi^4$. It is worth emphasizing that, within this theory, there is no need to introduce Lagrange multipliers to enforce the vanishing-torsion condition without trivializing the scalar field. Instead, this condition can be consistently implemented by setting $\lambda=1$.

The field equations are obtained by performing arbitrary variations of the total action Eq.\eqref{exact1}, with respect to the vierbein, Lorentz connection, the scalar field, and Maxwell field giving
\begin{subequations}\label{eom}
    \begin{align}
 0&=	\frac{1}{2}\epsilon_{abcd}R^{ab}\wedge e^c -\frac{\Lambda}{3!} \epsilon_{abcd}e^{a}\wedge e^{b}\wedge e^{c}-\kappa (*\tau^{(\phi)}_d+*\tau^{\text{(em)}}_d) \,,  \label{eomm}   \\
0&=		\epsilon_{abcd}T^{c}\wedge e^{d} -\kappa*\sigma_{ab}\,,\label{eomt} \\
0&= \epsilon_{abcd}\left[\lambda(2-\lambda) \text{D} Z^a \wedge e^b-\frac{1}{2}\phi R^{ab}- \lambda(3\lambda-5)Z^a\wedge T^b+\lambda \text{d}\phi \wedge \text{I}^a(T^b)\right]\wedge e^c \wedge e^d \notag\\
	&+\epsilon_{abcd}\left[\lambda\phi \text{D}(\text{I}^a T^b)\wedge e^c -2\lambda \phi\text{I}^a(T^b)\wedge T^c-\frac{1}{4}\frac{\text{d}V(\phi)}{d\phi} e^a \wedge e^b \wedge e^c\right]\wedge e^d  \,, \label{eomphi}\\
0&= \text{d}* F\,, \label{eomA}   
\end{align}
\end{subequations}
respectively. Additionally, we have defined the stress-energy and spin density three-forms as $\tau_a$ and $\sigma_{ab}$, respectively; they are
\begin{align}
*\tau^{(\phi)}_d &= \frac{1}{3}\epsilon_{abcd}\left(\frac{\phi^2}{4}R^{ab} + \lambda\phi Z^a T^b \right)\wedge e^c +\frac{1}{6}\epsilon_{abcd}\bigg[ \frac{\lambda(\lambda-2)}{2}Z^2 + V(\phi) \bigg] e^a\wedge e^b\wedge e^c \notag \\
 &- \frac{\lambda}{6}\epsilon_{abcd}\left[\phi\Diff{Z^a}+(3\lambda-5)Z^a\,\diff{\phi} + \phi Z_n\text{I}^nT^a \right]\wedge e^b\wedge e^c\,, \\
 *\tau^{(\text{em})}_d &=\frac{1}{8\pi}(F\wedge * (e_{d}\wedge F)-\text{I}_d F \wedge * F),\\
 *\sigma_{ab}  &=\frac{\left(  1-\lambda\right) }{6-\kappa\phi^{2}}\epsilon_{abcd}\, \phi\,\diff{\phi}\wedge e^{c}\wedge e^{d}\,.
\end{align}
The field equation for the spin connection can be solved algebraically, so that torsion is expressed directly in terms of the scalar field and its derivative as
\begin{equation}
T^{a}=\frac{\kappa\left(  1-\lambda\right)  }{\left( 6-\kappa
\phi^{2}\right)}\,\phi\,\diff{\phi}\wedge e^{a}\,.   \label{torsol}
\end{equation}
Equation \eqref{torsol} shows that the scalar field sources a highly constrained torsional sector. In fact, the torsion is entirely determined by the gradient of $\phi$ and is purely of vector-trace type, while its axial and traceless tensor irreducible components vanish identically. Therefore, the nonminimal conformal coupling does not excite the most general Riemann--Cartan torsional structure, but rather selects a restricted sector in which torsion is encoded in a single scalar degree of freedom. This feature considerably simplifies the geometric content of the solutions and makes transparent the role of the scalar field as the sole dynamical source of torsion in the model.
Thus, we conclude that the nonminimal coupling of the scalar field acts as a source of nontrivial torsion in this theory and activates the vector trace sector of the torsion through a single dynamical degree of freedom. The role of nonminimal couplings as sources of torsion has also been reported in Refs.~\cite{Toloza:2013wi,Cid:2017wtf,Barrientos:2019awg,Valdivia:2017sat}. Moreover, in the limit $\lambda \to 1$, torsion vanishes independently of the value of the scalar field. In contrast, for $\lambda \neq 1$, torsion remains nontrivial provided that the scalar field is not constant. In what follows, we solve the remaining field equations by adopting a static ansatz and show that the system admits black hole solutions endowed with nontrivial torsion.

\section{Exact solutions with nontrivial torsion and scalar field}\label{sec:solution}

We assume a static metric whose codimension-2 hypersurfaces of constant $t-r$ represent locally a constant curvature space. In particular, we consider 
\begin{align}\label{metric}
    \diff{s^2} = h(r)\left(-f(r)\diff{t^2} + \frac{\diff{r^2}}{f(r)} + r^2\diff{\Omega}^2 \right)\,,\;\;\;\;\; \mbox{where} \;\;\;\;\; \diff{\Omega}^2  = \text{d}\theta^2+\frac{1}{k}\sin^2(\sqrt{k}\theta)\text{d}\varphi^2
\end{align}
with Gaussian curvature $k=\pm 1$. For $k=1$, the line element of the base manifold corresponds to that of a two-sphere, whereas for $k=-1$, the base manifold corresponds to a hyperbolic surface. In the associated orthonormal frame, one has
\begin{equation}
     \text{e}^0=\sqrt{h(r)f(r)}\,\text{d}t\,,\;\;\text{e}^1=\sqrt{\frac{h(r)}{f(r)}}\,\text{d}r\,,\;\; \text{e}^2=r\sqrt{h(r)}\,\text{d}\theta\,,\;\;\text{e}^3=r\sqrt{\frac{h(r)}{k}}\sin(\sqrt{k}\theta)\,\text{d}\varphi,
\end{equation}

Additionally, the scalar field and electromagnetic field compatibles with the isometries of this metric depends on the radial coordinate only, namely, $\phi=\phi(r)$ and $A=A(r)\text{d}t$.

As we mentioned above, the field equation for the connection can be solved for the torsion in terms of the scalar field and derivatives thereof, whose solution is given in Eq.~\eqref{torsol}. This, in turn, implies that the functions $\omega_I(r)$, with $I=1,\ldots,8$, can be solved algebraically in terms of the scalar field and the metric functions. The nontrivial pieces of the connection are found to be
\begin{equation}
    \omega^{01}=\omega_{1}(r)\text{e}^0\,,\;\;\omega^{12}=\omega_{2}(r)\text{e}^2\,,\;\; \omega^{13}=\omega_{2}(r)\text{e}^3\,,\;\;\omega^{32}=\omega_{3}(r)\text{e}^3,
\end{equation}
where
\begin{subequations}\label{connection}
    \begin{align}
     \omega_1(r) &= \frac{1}{h(r)}\left[\sqrt{h(r)f(r)}\right]' + \frac{(1-\lambda)\sqrt{f(r)}\,\phi(r)\phi'(r)}{J(\phi)\sqrt{h(r)}}\,, \\
     \omega_2(r) &= - \frac{\sqrt{f(r)}}{2r^2h^{3/2}(r)}\left[h(r)\,r^2 \right]' - \frac{(1-\lambda)\sqrt{f(r)}\,\phi(r)\phi'(r)}{J(\phi)\sqrt{h(r)}}\,,\\
     \omega_{3}(r) &= \frac{1}{r}\sqrt{\frac{k}{h(r)}}\cot(\sqrt{k}\theta)
 \end{align}
\end{subequations}
with $J(\phi)=\phi^2 - \frac{3}{4\pi G}$, prime denotes differentiation with respect to the radial coordinate, i.e. $'=\diff{}/\diff{r}$, and the other components of the connection vanish on shell.

\subsection{Solutions with \texorpdfstring{$\Lambda = 0\, \text{and}\,  V(\phi)=0$} a}

Let us consider a spacetime with vanishing cosmological constant and a scalar field with a vanishing potential. We also assume a base manifold with positive curvature, $k=1$. In this case, the field equations for the vielbein and the scalar field admit solutions that can be written as
 \begin{subequations}\label{sol}
     \begin{align}
     f(r) &= \left(1-\frac{M\,G}{r} \right)^2, \label{EqBH} \\
     h(r) &= \frac{\left[(r-G M \left(1-Q\right))^{a}+(r-G M \left(1+Q\right))^{a}\right]^2}{4(r-G M)^2 \left[(r-G M \left(1-Q\right))(r-G M \left(1+Q\right))\right]^{a-1}}, \\
     \phi(r)&= \sqrt{\frac{3}{4\pi G}}\left[\frac{(r-G M \left(1-Q\right))^a-\left(r-G M \left(1+Q\right)\right)^{a}}{(r-G M \left(1-Q\right))^a+\left(r-G M \left(1+Q\right)\right)^{a}}\right]\,,\\
     A_t(r)&=-\frac{q}{r},
 \end{align}
 \end{subequations}
where we have defined $a \equiv  (2\lambda - \lambda^{2})^{-1/2}$, and where $M$ and $q$ are integration constants, related through
\begin{equation}
    Q^2=1-\frac{q^2}{GM^2}.
\end{equation}
The causal structure of these solutions depends on the values of the integration constants and on the parameter $a$. From Eq.~\eqref{EqBH}, we observe that, for $M>0$, the metric admits an event horizon at $r=GM$; therefore, the configuration describes a black hole. For $M=0$, the configuration does not admit a real scalar field. However, if the additional condition $q=0$ is imposed, the scalar field vanishes identically and the geometry reduces to flat spacetime. Finally, for $M<0$, the spacetime corresponds to a wormhole geometry.

It is worth emphasizing that these statements depend on the values of the parameter $a$ and the constant $q$, as will be demonstrated throughout this work.

On the other hand, we note that the solution in~\eqref{sol} is continuously connected to the configurations presented in~\cite{Bekenstein:1974sf, Bekenstein:1975ts} for the case $a=1$, provided that $M>0$. In this regime, the spacetime corresponds to a purely metric black hole, where the scalar field develops a singularity at the event horizon.

In the next section, the physical properties of the matter fields and the resulting causal structure associated with the different spacetime configurations obtained will be analyzed in detail.
\subsubsection{Global Properties of the Matter Fields and the Spacetime \label{sec:properties}}

To analyze the global properties of the solutions, we begin by examining the behavior of the scalar field. In particular, we require the scalar field to be nontrivial and to take real values for all positive values of the radial coordinate. This condition restricts the integration constants to satisfy $q^{2} < G M^{2}$, which in turn implies that $0 < Q \leq 1$. Moreover, the parameter $a>1$ must take integer values.

For $M>0$, if $a$ is an even integer, the scalar field remains bounded within the interval $-\sqrt{\frac{3}{4 \pi G}} \leq \phi \leq \sqrt{\frac{3}{4 \pi G}}$ and is regular everywhere. By contrast, if $a$ is an odd integer, the scalar field develops a singularity at the point $r=GM$.

On the other hand, for $M<0$ and for any value of $a$, the scalar field remains bounded within the interval $-\sqrt{\frac{3}{4 \pi G}} \leq \phi < 0$ and is regular throughout the spacetime.

We proceed to analyze the main properties of the spacetime. In the asymptotic region $r \to \infty$, the metric components take the form
\begin{eqnarray}
    g_{tt}&=&-h(r)f(r)\approx -1+\frac{2GM}{r}+\mathcal{O}(r^{-2}),\\
    g^{rr}&=&\frac{f(r)}{h(r)}\approx 1-\frac{2GM}{r}+\mathcal{O}(r^{-2}).
\end{eqnarray}
For the angular components of the metric, the leading-order behavior is $\mathcal{O}(r^{2})$; consequently, the metric approaches the flat spacetime form in the asymptotic region. On the other hand, a direct computation of the Kretschmann invariant shows that it is proportional to
\begin{equation}
    K \propto \frac{(r-GM)^6\left[(r-G M \left(1-Q\right))(r-G M \left(1+Q\right))\right]^{2a-6}}{r^{8}}.
\end{equation}
Therefore, we observe that the spacetime may exhibit singularities whose existence depends on the value of the parameter $a$. On the other hand, the solutions possess nontrivial torsion; consequently, torsional invariants must also be analyzed. In particular, we consider $T = T_{\mu\nu\rho}T^{\mu\nu\rho}$, which is given by
\begin{equation*}
    T=\frac{24(a^2-1)(GMQ)^{2}(r-MG)^{4} \left[(r-G M \left(1+Q\right))^{a}-(r-G M \left(1-Q\right))^{a}\right]^2 }{r^{2}\left[(r-G M \left(1-Q\right))(r-G M \left(1+Q\right))\right]^{3-a}\left[(r-G M \left(1+Q\right))^{a}+(r-G M \left(1-Q\right))^{a}\right]^4}.
\end{equation*}

Using these geometric quantities, together with the properties of the scalar field, we analyze the global properties of the solutions according to the different possible values of the integration constants and the parameter $a$, as summarized in Table~\ref{tab:flat-compact}.

For $M<0$ and $0<Q\leq1$:

\begin{itemize}
\item For any $a>1$, the scalar field is regular throughout the spacetime. Moreover, it attains its minimum value at $r=0$ and its maximum value in the limit $r \to \infty$. If $a<5$, the spacetime possesses a naked singularity at $r=0$. However, if $Q=1$ and $a \geq 5$, the curvature and torsional singularities are removed. By introducing the areal radius $R(r)=r \sqrt{h(r)}$, one observes that spacetime exhibits two asymptotic regions, since $R(r \to \infty) \to \infty$ and $R(r \to 0) \to \infty$. In addition, this function admits a minimum between both regions and therefore defines a throat. One can also verify that no horizons are present at this throat, since $f(r)>0$ for all $r$, which implies that $g_{tt}$ does not vanish for $r>0$. Consequently, the spacetime can be geometrically interpreted as a traversable wormhole with torsion. We emphasize that the traversability of the wormhole solutions discussed here is established at the level of the metric geometry. More precisely, our characterization is based on the existence of a throat connecting two asymptotic regions and on the absence of event horizons, as inferred from the behavior of the areal radius and the causal structure determined by the metric. In this sense, the term traversable refers to the possibility of causal communication through the throat in the metric spacetime. A more detailed analysis of particle motion based on autoparallel curves of the full torsionful connection, and its comparison with the corresponding metric geodesics, lies beyond the scope of the present work and will be addressed elsewhere.
\end{itemize}

For $M>0$ and $0<Q\leq1$:

\begin{itemize}
\item If $a=2n-1$\footnote{Imposing the condition $\lambda\in[0,1]$ leads to 
$\lambda=1-\frac{2}{2n-1}\sqrt{n(n-1)}$.} with $n\in \mathbb{N}$, the scalar field is regular everywhere except at the point $r=GM$. Moreover, it attains its maximum and minimum values at $r_{\pm}=GM(1\pm Q)$, respectively. In the case $n=1$, torsion vanishes and the spacetime reduces to the Bekenstein solution. For $n\geq 2$, the spacetime possesses a singularity at $r=0$ covered by an event horizon located at $r=GM\equiv r_{h}$, since this hypersurface is null and corresponds to a Killing horizon, that is, $g^{rr}(r_{h})=g_{tt}(r_{h})=0$. Therefore, this configuration represents a torsional extension of the Bekenstein solution describing a black hole with a regular event horizon and nontrivial torsion. Finally, if $Q=1$ ($q=0$) and $n>2$, the curvature and torsional singularities at $r=0$ can be removed, and the resulting configuration corresponds to a regular black hole~\cite{Ayon-Beato:1998hmi} endowed with nontrivial torsion.

\item If $a=2n$ \footnote{Imposing the condition $\lambda\in[0,1]$ leads to 
$\lambda=1-\frac{1}{2n}\sqrt{4n^2-1}$.} with $n\in \mathbb{N}$, the scalar field is regular throughout the spacetime; moreover, it vanishes at $r=GM$ and in the limit $r\to \infty$. It also attains its maximum and minimum values at $r_{\pm}=GM(1\pm Q)$, respectively. In the case $n=1$, the spacetime possesses three naked curvature and torsional singularities located at $r=0$ and $r=r_{\pm}$. If $n\geq 2$, the singularities at $r=r_{\pm}$ disappear and, consequently, only the central singularity remains. On the other hand, if $Q=1$ and $n>2$, this singularity can also be removed. However, the spacetime undergoes a change of signature in the region $r<r_{+}$. Therefore, this solution corresponds to the exterior region of an extremal Reissner–Nordström geometry with nonvanishing torsion and a nontrivial scalar field, for $r\geq r_{+}$, with $n\geq2$ and $0<Q<1$.
\end{itemize}

Finally, we note that the first-order formulation, by treating the metric and the connection as independent fields, allows for a one-parameter extension of conformal symmetry. This generalization gives rise to new spacetime configurations with nontrivial torsion, which are continuously connected to purely metric solutions in the limit where the standard conformal symmetry is recovered. In what follows, we shall study exact solutions that are asymptotically locally AdS in the presence of a nontrivial scalar potential.
\begin{table}[!ht]
	\centering
	\caption{Asymptotically flat branches $(\Lambda=0,\;V(\phi)=0)$.}
	\label{tab:flat-compact}
	\renewcommand{\arraystretch}{1.18}
	\small
	\begin{tabular}{||c|c|c|c|c||}
		\hline\hline
		\textbf{Branch} & \boldmath{$a$} & \boldmath{$\phi$} & \textbf{Horizon} & \textbf{Geometry} \\
		\hline\hline
		$M<0$ 
		& $1<a<5$ 
		& regular 
		& none 
		& Naked singularities \\
		\hline

		$M<0$, $q=0$ 
		& $a\geq 5$ 
		& regular 
		& none 
		& Torsional traversable wormhole \\
		\hline

		$M>0$ 
		& $a$ odd, $a\geq 3$ 
		& singular at $r=GM$ 
		& $r_h=GM$ 
		& Torsional Bekenstein BH \\
		\hline

		$M>0$, $q=0$ 
		& $a$ odd, $a\geq 5$  
		& singular at $r=GM$ 
		& $r_h=GM$ 
		& Torsional regular BH \\
		\hline

		$M>0$ 
		& $a$ even, $a\geq 4$ 
		& regular 
		& none 
		& Exterior region of extremal R--N \\
		\hline

		$M=0,\ q=0$ 
		& -- 
		& $\phi=0$ 
		& none 
		& Minkowski \\
		\hline\hline
	\end{tabular}
\end{table}

\subsection{Solutions with \texorpdfstring{$\Lambda = -\frac{3}{\ell^2}\, \text{and}\,  V(\phi)\neq 0$} a}

In this section, we investigate the space of solutions of the theory by solving the field equations~\eqref{eom} in the presence of a negative cosmological constant. To this end, we consider a one-parameter potential given by
\begin{equation}\label{Vphi}
V(\phi)=\frac{\Lambda\pi G J^2(\phi)}{9} \left[ \left(\frac{3+\sqrt{12\pi G}\phi}{3-\sqrt{12\pi G}\phi}\right)^{-\frac{1}{2a}}+\left(\frac{3+\sqrt{12\pi G}\phi}{3-\sqrt{12\pi G}\phi}\right)^{\frac{1}{2a}} \right]^{4}
-\frac{\Lambda}{8\pi G},
\end{equation}
where $J(\phi)=\phi^2 - \frac{3}{4\pi G}$. The vanishing potential is recovered in the vanishing-torsion limit, namely $a \to 1$ ($\lambda \to 1$). It is worth noting that Eq.~\eqref{Vphi} possesses extrema at $\phi_0 = 0$ and $\phi_0 = \pm\sqrt{\frac{3}{4\pi G}}$. As will be shown below, these values arise as limiting configurations of the scalar field. The stability of these extrema depends on the value of the cosmological constant evaluated at these points.

The potential satisfies $V''(0)=-\tfrac{2\Lambda}{3}(\lambda-1)^2$.\footnote{For $\lambda\neq 1$, this potential contributes to the mass term of the scalar field.} Hence, for $\Lambda<0$, the extremum at $\phi_0=0$ corresponds to a global minimum. Finally, for $\lambda=1$, torsion vanishes and the results of Ref.~\cite{Martinez:2004nb} are recovered.

To solve Eqs.~\eqref{eom}, we consider a metric whose transverse section has negative curvature. Moreover, the nontrivial components of the spin connection retain the structure given in~\eqref{connection}, now evaluated for $k=-1$. The field equations for the vierbein and the scalar field are then solved by
\begin{subequations}\label{sol2}
     \begin{align}
     f(r) &= \frac{r^2}{\ell^2}-1 , \label{EqBH1} \\
     h(r) &= \frac{\left[(r+\sqrt{G} \,q)^{a}+(r-\sqrt{G} \,q)^{a}\right]^2}{4r^2 \left[(r+\sqrt{G} \,q)(r-\sqrt{G} \,q)\right]^{a-1}}, \\
     \phi(r)&= \sqrt{\frac{3}{4\pi G}}\left[\frac{(r+\sqrt{G} \,q)^{a}-(r-\sqrt{G} \,q)^{a}}{(r+\sqrt{G} \,q)^{a}+(r-\sqrt{G} \,q)^{a}}\right]\,,\\
     A_t(r)&=-\frac{q}{r},
 \end{align}
 \end{subequations}
where $q$ is an integration constant. From this expression, we observe that the nontrivial nature of the scalar field arises from the presence of the electromagnetic field, since for $q=0$ one obtains $\phi(r)=0$, and the spacetime corresponds to a topological black hole.

In the vanishing-torsion case, a massless black hole solution is recovered whose energy--momentum tensor vanishes despite the presence of nontrivial matter fields; that is, the configuration corresponds to a stealth solution~\cite{Martinez:2005di}. In what follows, we explore the properties of this solution in the presence of nontrivial torsion.

 \subsubsection{Global Properties of the Matter Fields and the Spacetime\label{sec:properties11111111}}

 In order to characterize the global behavior of the solutions, we first examine the properties of the scalar field. We demand that the scalar field be nontrivial and real for all positive values of the radial coordinate. This requirement restricts the parameter $a>1$ to take integer values. Furthermore, since the solution reported in~\cite{Martinez:2005di} belongs to this class of configurations, we additionally impose the condition $q^{2}<\ell^{2}/G$, which guarantees a positive entropy.

Under these assumptions, when $a$ is an even integer, the scalar field remains finite within the range $-\sqrt{\frac{3}{4 \pi G}} \leq \phi \leq \sqrt{\frac{3}{4 \pi G}}$ and is well behaved throughout the spacetime. In contrast, for odd integer values of $a$, the scalar field becomes singular at the origin, $r=0$.

We now turn to the analysis of the geometric properties of the spacetime. In the asymptotic regime $r \to \infty$, the metric components take the form
\begin{eqnarray}
    g_{tt}&=&-h(r)f(r)\approx \frac{\Lambda r^{2}}{3}+1+ \frac{\Lambda (a^{2}-1) G q^{2}}{3} +\mathcal{O}(r^{-2}),\\
    g^{rr}&=&\frac{f(r)}{h(r)}\approx -\frac{\Lambda r^{2}}{3}-1+ \frac{\Lambda (a^{2}-1) G q^{2}}{3}+\mathcal{O}(r^{-2}).
\end{eqnarray}
The metric corresponds to an asymptotically locally AdS spacetime. We note that the presence of torsion induces an effective curvature on the transverse section, which can be inferred from the leading constant term in the $g_{tt}$ component. On the other hand, the Kretschmann invariant can be computed and is found to be proportional to
\begin{equation}
    K \propto \frac{\left[(r+\sqrt{G} \,q)(r-\sqrt{G} \,q)\right]^{2a-6}}{\left[(r+\sqrt{G} \,q)^a+(r-\sqrt{G} \,q)^a\right]^8}.
\end{equation}
Therefore, the spacetime may exhibit singularities whose existence depends on the value of the parameter $a$. The torsional invariant $T$ is given by
\begin{equation*}
    T=\frac{24(a^2-1)Gq^{2}r^2 (r^2-\ell^2)(r^2-G\,q^2)^{a-3
    } \left[(r-\sqrt{G}\,q)^{a}-(r+\sqrt{G}\, q)^{a}\right]^2 }{\ell^2\left[(r-\sqrt{G}\,q)^{a}+(r+\sqrt{G}\, q)^{a}\right]^4}.
\end{equation*}
On the basis of these geometric quantities and the properties of the scalar field, we proceed to analyze the global properties of the solutions under the condition $q^{2}<\ell^{2}/G$. The corresponding classification according to the different possible values of the parameter $a$ is summarized in Table~\ref{tab:ads-compact}.

\begin{itemize}
\item If $a=2n$ with $n\in \mathbb{N}$, the scalar field is regular everywhere. Moreover, it attains its maximum and minimum values at $r_{\pm}= \pm \sqrt{G}\,q$ (depending on the sign of $q$), respectively. The spacetime also possesses an event horizon at $r=\ell$ for all values of $n$. For $n=1$, a curvature and torsional singularity appears at the point $r_{+}$ or $r_{-}$, which is covered by the horizon. Therefore, the solution describes a charged torsional black hole dressed by a nontrivial scalar field that remains regular everywhere. If $n>1$, these singularities can be removed, and the point $r_{+}$ or $r_{-}$ becomes an asymptotic region of the spacetime.

\item If $a=2n-1$ with $n\in \mathbb{N}$, the scalar field is regular everywhere except at $r=0$. Moreover, it attains its maximum and minimum values at $r_{\pm}= \pm \sqrt{G}\,q$ (depending on the sign of $q$), respectively. The spacetime also possesses an event horizon at $r=\ell$ for all values of $n$. For $n=1$, torsion becomes trivial and the spacetime corresponds to a topological stealth black hole. If $n\geq2$, the configuration exhibits a curvature and torsional singularity at $r=0$. Therefore, the solution describes a torsional black hole whose geometric quantities and matter fields diverge at $r=0$, behind the event horizon.
\end{itemize}

From these results, it is natural to restrict the parameter space to $q^{2}<\ell^{2}/G$, since in this case the curvature and torsional singularities remain hidden behind an event horizon. In contrast, when $q^{2} \geq \ell^{2}/G$, the entropy becomes negative for $a=1$, whereas for $a>1$ the spacetime develops naked singularities.

\begin{table}[!ht]
	\centering
	\caption{Asymptotically locally AdS branches $\left(\Lambda=-3/\ell^2,\;V(\phi)\neq 0\right)$.}
	\label{tab:ads-compact}
	\renewcommand{\arraystretch}{1.18}
	\small
	\begin{tabular}{||c|c|c|c|c||}
		\hline\hline
		\textbf{Branch} 
		& \boldmath{$a$} 
		& \boldmath{$\phi$} 
		& \textbf{Horizon} 
		& \textbf{Interpretation} \\
		\hline\hline

		$q\neq 0$ 
		& $a$ even 
		& regular 
		& $r=\ell$ 
		& Torsional charged topological BH \\
		\hline

		$q\neq 0$ 
		& $a$ odd 
		& singular at $r=0$ 
		& $r=\ell$ 
		& Torsional charged topological BH \\
		\hline

		$q\neq 0$ 
		& $a=1$  
		& singular at $r=0$ 
		& $r=\ell$ 
		& Torsionless stealth BH \\
		\hline

		$q=0$ 
		& -- 
		& $\phi=0$ 
		& $r=\ell$ 
		& Topological BH \\
		\hline\hline
	\end{tabular}
\end{table}

\newpage 
\section{Discussion\label{sec:discussion}}

In this work, we introduced a one-parameter generalization of conformal transformations within a gravitational theory formulated in the first-order formalism, in which the metric and connection are treated as independent dynamical fields. This extension makes it possible to implement the conformal coupling of the scalar field directly in the first-order framework while retaining its standard conformal weight. The resulting model naturally accommodates nontrivial torsion and includes, as a particular limit, the conventional conformal coupling of the scalar field, recovered for a specific value of the parameter. In this limit, torsion vanishes without requiring the scalar field itself to become trivial. Building upon this framework, we coupled the theory to Einstein–Cartan gravity and to an electromagnetic field, and investigated exact static solutions in asymptotically flat and asymptotically AdS settings, as well as in scenarios where conformal symmetry is explicitly broken through the introduction of a scalar potential.

In order to construct exact solutions, we adopted a static ansatz whose transverse section may be given by either a sphere or a hyperboloid. For the matter fields, and in accordance with the spacetime isometries, we assumed purely radial dependence. Depending on the values of the integration constants and of the parameter that generalizes the conformal transformation, we obtained several spacetime configurations with nontrivial torsion. In the asymptotically flat regime, we found a novel wormhole solution in Einstein–Cartan gravity with a scalar field that remains regular throughout the spacetime, as well as a torsional extension of the Bekenstein black hole, which in the absence of electric charge reduces to a regular black hole. On the other hand, in the asymptotically AdS case and in the presence of a scalar potential, we identified topological black hole solutions with torsion for different values of the parameter $a$, in which the scalar field remains regular everywhere. The very existence of these black holes is intrinsically linked to the presence of electric charge. Finally, in the vanishing-torsion limit, the previously known black hole solutions reported in the literature for purely metric theories are recovered.

The results obtained here open several avenues for the study of phenomenological aspects associated with spacetimes endowed with nontrivial torsion. Among these, it is particularly relevant to analyze the energy conditions within the framework of Einstein–Cartan gravity for these new solutions~\cite{DiGrezia:2017daq,DeBenedictis:2018mmx}, as well as to contrast their behavior with that of analogous configurations formulated in general relativity. Likewise, it would be of interest to study autoparallel trajectories and to examine singularity theorems in the presence of torsion for these configurations~\cite{Puetzfeld:2013sca,Luz:2017ldh,Luz:2019kmm,Obukhov:2021uor,vandeVenn:2024ayq}. Furthermore, the analysis of the thermodynamic properties of the black hole solutions could contribute to a deeper understanding of the modifications induced by torsion \cite{Blagojevic:2006jk,Blagojevic:2006hh,Blagojevic:2022etm,Aviles:2023igk,Aviles:2025nah}. 
Another natural direction would be to investigate whether the torsional hair present in these geometries leaves any imprint on perturbative observables, such as wave propagation, in analogy with studies of gravitational-wave propagation in Einstein–Cartan backgrounds sourced by spin densities~\cite{Elizalde:2022vvc}.
Finally, it would also be worthwhile to explore cosmological scenarios in which spacetime torsion plays a significant role~\cite{Galiakhmetov:2012cp,Luz:2023uhy,Alam:2025qkx}. These questions remain open and constitute natural directions for future research.

\begin{acknowledgments}
We thank Eloy Ayón-Beato, Cristóbal Corral, Borja Diez, Oscar Fuentealba and David Tempo for insightful comments and remarks. The work of L.A. is partially supported by Agencia Nacional de Investigaci\'{o}n y Desarrollo (ANID) through Fondecyt Iniciación grants N$^{\mathrm{o}}$11261098 and SIA-ANID grant N$^{\mathrm{o}}$85220027. O.V. acknowledges Fondecyt regular grant N$^{\mathrm{o}}$1251523.
	\end{acknowledgments}

\appendix
 
\section{Action and field equations in tensor components\label{sec:appendix}}

For the sake of comparison, here we provide the action and field equations in tensor components to perform the analysis on a coordinate basis. In this case, the action principle in Eq.~\eqref{exact1} can be rewritten as
\begin{align}
I  & =\int_{\mathcal{M}_{4}}\diff{^4}x\sqrt{|g|}\left[  \frac{1}{2\kappa}\left(  R-2\Lambda\right)-\frac{1}{16\pi}F_{\mu \nu}F^{\mu \nu}
-\frac{1}{12}\phi^{2}R-\lambda\left[  1-\frac{\lambda}{2}\right]
(\nabla\phi)^2-\frac{\lambda}{3}\phi T_\lambda\nabla^{\lambda
}\phi -V\left(  \phi\right)  \right]\,, 
%& -\partial_{\mu}\left(  \frac{\lambda}{2}\phi\partial^{\mu}\phi\sqrt{|g|}\right)  d^{4}x
\end{align}
where $T_\lambda := T^{\sigma}{}_{\lambda\sigma}$ is the trace of the torsion tensor. The field equation for the vierbein can be written as
\begin{equation}
G_{\mu\nu}+g_{\mu\nu}\Lambda=\kappa\, (\tau^{(\phi)}_{\mu\nu}+\tau^{(em)}_{\mu\nu})\,,\label{einstt}
\end{equation}
where $G_{\mu\nu}$ denotes the Einstein tensor constructed out of the torsionful connection and
\begin{align}
\tau^{(\phi)}_{\mu\nu}  & =\frac{1}{6}\phi^{2}G_{\mu\nu}-g_{\mu\nu}V\left(  \phi\right)
-\frac{1}{2}\lambda\left(  \lambda-2\right)  g_{\mu\nu}(\nabla\phi)^2+\frac{1}{3}\lambda\phi\nabla_{\mu}\phi
T_\nu \nonumber \\
& +\frac{1}{3}\lambda\left(  3\lambda-5\right)\left[  g_{\mu\nu}%
(\nabla\phi)^2-\nabla_{\mu
}\phi\nabla_{\nu}\phi\right]  +\frac{1}{3}\lambda\phi\left(  g_{\mu\nu}\Box\phi-\nabla_{\nu}\nabla_{\mu}\phi\right) \,,\\
\tau^{(em)}_{\mu\nu}  &= -\frac{1}{4\pi}\left(F_{\mu \alpha}{F^{\alpha}}_{\nu}+\frac{1}{4}g_{\mu\nu}F_{\alpha \beta}F^{\alpha \beta}\right)
\end{align}
are the stress-energy tensors for the conformally-coupled scalar and electromagnetic field, respectively. Here, $\Box=\nabla_\mu\nabla^\mu$ is constructed out of the torsionful covariant derivative. It is worth noting that the field equation Eq.~\eqref{einstt} is not symmetric in general since $\left[\nabla_\mu,\nabla_\nu \right]\phi =  T^{\lambda}{}_{\mu\nu}\nabla_\lambda\phi$. The skew-symmetric piece arises from the fact that the vielbein is not necessarily symmetric in its two indices; the antisymmetric components are related to the presence of torsion. 

The field equation for the scalar field is given by
\begin{equation}
2\lambda\left(
1-\frac{\lambda}{2}\right)  \Box\phi-\frac{1}{6}\phi R-\frac{\partial V}{\partial\phi}-\lambda\left(
\lambda-2\right)  T^\mu\nabla_\mu\phi+\frac{1}{3}\lambda
\phi\nabla_\mu T^\mu+\frac{1}{3}\lambda\phi
T_\mu T^\mu=0\,,
\end{equation}
Finally, the field equation for the spin connection is given by
\begin{equation}
T_{~\alpha\beta}^{\mu} + 2\delta^\mu_{[\alpha}T_{\beta]}=-\frac{2\kappa\left(
\lambda-1\right)  }{\left(6-\kappa \phi^{2}\right)}\delta^\mu_{[\alpha}\nabla_{\beta]}\phi^2\,,
\end{equation}
whose solution is
\begin{equation}
T_{~\mu\nu}^{\rho}= \frac{\kappa\left(\lambda-1\right)  }{\left(
6-\kappa\phi^{2}\right)  }\,\delta^\rho_{[\mu}\nabla_{\nu]}\phi^2 \,.%=\frac{\kappa\left(  1-\lambda\right)  }{\left(6-\kappa\phi^{2}\right)  }\phi\left(  \mathrm{\partial}_{\mu}\phi\delta_{\nu}^{\rho}-\mathrm{\partial}_{\nu}\phi\delta_{\mu}^{\rho}\right) 
\end{equation}

\bibliography{GB2019}

\end{document}